\documentclass[sigconf]{acmart}

\AtBeginDocument{%
  \providecommand\BibTeX{{%
    \normalfont B\kern-0.5em{\scshape i\kern-0.25em b}\kern-0.8em\TeX}}}

\setcopyright{acmcopyright}
\copyrightyear{2023} 
\acmYear{2023} 
\setcopyright{rightsretained} 
\acmConference[SIGMOD-Companion '23]{Companion of the 2023 International Conference on Management of Data}{June 18--23, 2023}{Seattle, WA, USA}
\acmBooktitle{Companion of the 2023 International Conference on Management of Data (SIGMOD-Companion '23), June 18--23, 2023, Seattle, WA, USA}
\acmDOI{10.1145/3555041.3589399}
\acmISBN{978-1-4503-9507-6/23/06}

\usepackage{graphicx}
\usepackage[utf8]{inputenc} 
\usepackage{amsmath,amsthm} 
\usepackage{url}
\usepackage{fancyvrb}
\usepackage{wrapfig}
\usepackage{xspace}
\usepackage{subcaption}
\usepackage{booktabs}
\usepackage{hyperref}
\usepackage{tikz-cd}
\usepackage{tikz}
\usepackage{neuralnetwork}
\hypersetup{
colorlinks=false,
citebordercolor={green}
}

\newcommand{\sys}{\textsc{Dangoron}\xspace}
\newcommand{\bm}{\textsc{Tomborg}\xspace}

\newcommand{\eat}[1]{}

\theoremstyle{plain}

\usepackage{etoolbox,xspace}
\usepackage{algorithmicx}
\usepackage{algorithm}
\usepackage{algpseudocode}
\algtext*{EndWhile}
\algtext*{EndIf}
\algtext*{EndFunction}
\algtext*{EndFor}
\algtext*{EndElse}
\algrenewcommand\algorithmicrequire{\textbf{Input:}}
\algrenewcommand\algorithmicensure{\textbf{Output:}}

\newcommand{\squishlist}{
 \begin{list}{$\bullet$}
  { \setlength{\itemsep}{0pt}
     \setlength{\parsep}{1pt}
     \setlength{\topsep}{1pt}
     \setlength{\partopsep}{0pt}
     \setlength{\leftmargin}{1em}
     \setlength{\labelwidth}{1em}
     \setlength{\labelsep}{0.5em} } }

\newcommand{\squishend}{\end{list}
}

\begin{document}

\title{\sys: Network Construction on Large-scale Time Series Data across Sliding Windows}

\author{Yunlong Xu}
\orcid{0000-0003-2589-7232}
\affiliation{
  \institution{}
  \streetaddress{}
  \city{University of Rochester}
  \state{}
  \country{}
  \postcode{}
}
\email{yxu103@u.rochester.edu}

\author{Peizhen Yang}
\orcid{}
\affiliation{
  \institution{}
  \streetaddress{}
  \city{University of Rochester}
  \state{}
  \country{}
  \postcode{}
}
\email{pyang11@u.rochester.edu}

\author{Zhengbin Tao}
\affiliation{%
  \institution{University of Rochester}
  \streetaddress{}
  \city{}
  \country{}}
  \email{ztao6@UR.Rochester.edu}

\renewcommand{\shortauthors}{Xu, et al.}

\renewcommand{\shorttitle}{\sys: Network Construction on Large-scale Time Series Data across Sliding Windows}



\maketitle

\section{Problem and Motivation}
\label{sec:problem}
The monitoring and analysis of large-scale time series data has become a significant area of interest within the database community. Constructing dynamic correlation-based complex networks is a highly effective method for extracting insights from extensive time series data across a range of disciplines, such as neuroscience~\cite{hutchison2013dynamic,cabral2017functional,thirion2014fmri}, climate science~\cite{kim2019complex,gozolchiani2008pattern}, and finance~\cite{tilfani2021dynamic,kenett2010dynamics}. In a correlation-based network, nodes are defined by their respective time series, while edges represent the correlations between nodes within a specified time range. Typically, in most analyses, a series of correlation matrices are generated by calculating the pairwise correlation of all time series within sliding windows, given a query time range, a query window size, and a sliding step. Pearson's correlation serves as one of the most prevalent measures across various domains~\cite{jayaweera2018reliability, krendl2022social, betzel2022network}.

\begin{figure}
    \vspace{-0.3cm}
    \centering
    \includegraphics[width=0.99\linewidth]{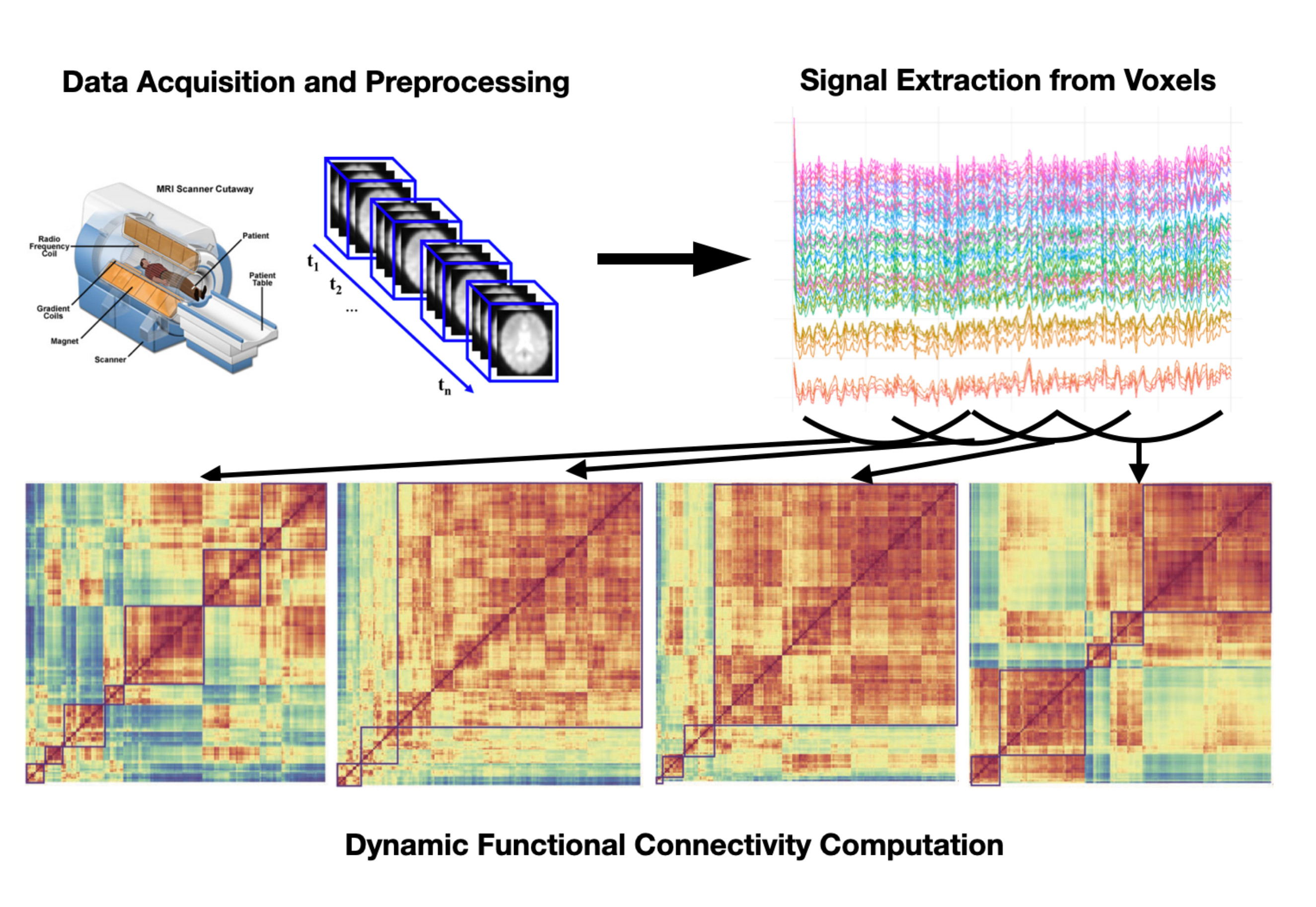}
    \caption{Network Construction on FMRI across Sliding Windows} \vspace{-12pt}
    \label{fig:fc_comp}
\end{figure}

 \textbf{Motivation Example.} Recently, dynamic analysis of blood oxygenation level-dependent functional magnetic resonance imaging (BOLD-fMRI) has provided valuable insights into brain network properties by quantifying functional connectivity metric changes over time~\cite{hutchison2013dynamic}. This analysis relies on 4-D fMRI data, which contains temporal BOLD activity recordings~\cite{cabral2017functional}. Each 3-D fMRI dataset comprises approximately 100K-10M regions at each time point. Traditional 4-D fMRI data analysis involves brain parcellation on region-based connectivity matrices, which refers to distinct, discontinuous yet closely interacting brain partitions~\cite{thirion2014fmri}. A superior alternative includes building a graph on voxel-level data and performing feature selection and graph embedding. However, constructing a network with raw voxel-level data is computationally infeasible due to the time-consuming pairwise correlation calculations required for voxel-based functional connectivity matrices within sliding windows. This paper introduces techniques to compute a sequence of correlation matrices on 3-D fMRI data for all brain regions

\textbf{Problem Definition.}
We are given a collection of time series represented by matrix $\mathcal{X}$ of size $N \times L$. Here, we have $N$ time series, and each row represents a time series of length $L$. Each element $x_{ij}$ in $\mathcal{X}$ denotes the time-stamped value of a variable collected at location $i$ at time $j$. We use $X_i$ to represent the $i$th row, which corresponds to the $i$th time series. We assume all time series in $\mathcal{X}$ are synchronized, meaning each time series has a value available at every periodic time interval or time resolution. This can be achieved through aggregation and interpolation on non-synchronized series. At query time, a user defines the query range $r=(s,e)$, where $s$ and $e$ are the starting and ending points, respectively, the query window size $l$, the sliding step size $\eta$, and a threshold $\beta$. The problem requires computing a series of correlation matrices, $\mathcal{C}={C^0, C^1, \cdots, C^\gamma}$, where each $C^k$ is the correlation matrix of $\mathcal{X}{k*\eta:k*\eta+l}$ (the submatrix of $\mathcal{X}$ from the $k*\eta$th column to the $k*\eta+l$th column). In $C^k$, each element $c{ij}^k(\geq \beta)$ denotes the Pearson correlation between $X_i$ and $X_j$ within the range of the $k$th sliding window. If $c_{ij}^k < \beta$, we replace it with $0$.

The core task involves addressing the problem of large-scale, all-pair time-series correlation calculation across sliding windows. The key challenges include: 1) ensuring the efficiency of network construction and updates for large-scale data to achieve interactivity, 2) maintaining the robustness of the methods on datasets with varying distributions, and 3) optimizing the effectiveness of sketching techniques for subsequent network analysis.

\section{Background and Related Work}
\label{sec:bg}

Correlation calculation across sliding windows has been explored through frequency-based transform~\cite{zhong2020filcorr,sakurai2005braid,zhu2002statstream,mueen2010fast} and random projection~\cite{yagoubi2018parcorr,zhang2009adaptive} approaches. Both aim to reduce time series dimensionality and identify highly-correlated pairs in low-dimensional spaces. Basic windows~\cite{zhu2002statstream} were first proposed for time-series pairs with high correlation, later improved in parallel~\cite{mueen2010fast}. Parcorr~\cite{yagoubi2018parcorr} optimized incremental computation, achieving state-of-the-art efficiency. TSUBASA~\cite{xu2022tsubasa}, the latest work, computes exact pairwise correlations on arbitrary time windows using a novel sketch framework, but lacks efficiency for sliding queries.

Existing techniques, except TSUBASA, face robustness issues due to data-dependency. Frequency-based transform methods struggle with dimensionality reduction for certain time series, only succeeding when energy concentrates in a few domains~\cite{zhang2009adaptive}. To date, no benchmark has been proposed to systematically test solution robustness.

\section{Approach and Uniqueness}
\label{sec:approach}

In this work, we introduce \sys, a framework for efficiently computing dynamic correlation matrices with high accuracy, and \bm, a benchmark for generating time series datasets to test framework robustness.

The core concept of \sys is the relatively stable correlation when transitioning to the next sliding window. By computing future upper and lower bounds of correlation, we can optimize calculations. For instance, if the current correlation is below the threshold and the next upper bound is also below the threshold, we can skip the next time step computation. Estimating bounds for the next $k$th window allows us to determine the number of skippable steps and directly jump to the desired window.
\begin{figure}[!ht]
  \centering 
  \includegraphics[width=0.38\textwidth]{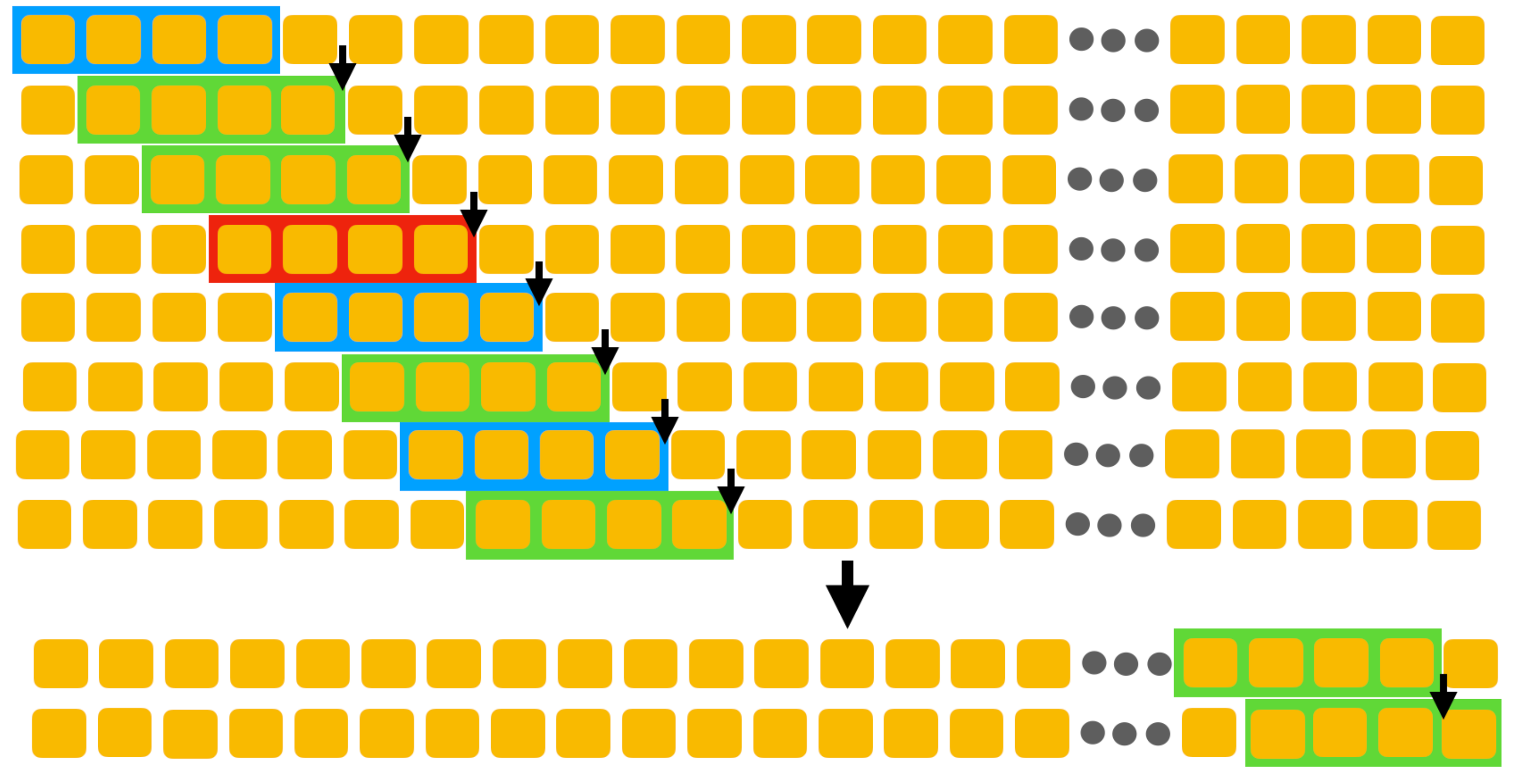} \vspace{-12pt}
  \caption{Jumping Structure of \sys} 
  \label{dangoron} 
\end{figure}

We employ the basic window framework by dividing a series into smaller basic windows, enabling data processing in smaller batches. Referring to the \textbf{Problem Definition}, let's consider a $C_k$ from $\mathcal{C}$ and denote it as $C$. We select the $i$th and $j$th rows from $\mathcal{X}$ within the time range specified by $C$, and label them as $x = [{\textbf x}_1, \ldots, {\textbf x}_m]$ and $y = [{\textbf y}_1, \ldots, {\textbf y}_m]$. We define the sizes of basic windows as $\textbf{B}=[B_1,B_2,\ldots , B_m]$, where $B_i$ represents the size of the $i$-th basic window. The exact Pearson's correlation between $x$ and $y$ is given by:

\newcommand\numberthis{\addtocounter{equation}{1}\tag{\theequation}}
\begin{align*}
  \vspace{-0.8cm}
  \numberthis \label{eq:exactcorr}Corr&(x,y)=\frac{\sum_{j=1}^{n_s}B_j(\sigma_{x_j} \sigma_{y_j} c_j + \delta_{x_j}\delta_{y_j})}{\sqrt{\sum_{i=1}^{n_s}B_i({\sigma_{x_i}}^{2}+{\delta_{x_i}^2})} \sqrt{\sum_{i=1}^{n_s}B_i({\sigma_{y_i}}^{2}+{\delta_{y_i}^2})}}\\
  &\delta_{x_i}=\overline{x_i}-\frac{\sum_{k=1}^{n_s}\overline{x_k}}{n_s},~~\delta_{y_i}=\overline{y_i}-\frac{\sum_{k=1}^{n_s}\overline{y_k}}{n_s}
\end{align*}
 $\sigma_{x_i}$ ($\sigma_{y_i}$) represents the standard deviation of the basic window for $x_i$ ($y_i$), $c_{i}$ is the correlation of basic windows $x_i$ and $y_i$, $\overline{x_i}$ ($\overline{y_i}$) is the mean of the basic window $x_i$ ($y_i$), and $n_s$ denotes the number of basic windows in a query window. Using Equation~\ref{eq:exactcorr}, we can pre-compute and store basic window statistics and calculate correlations for arbitrary query windows and sizes.

Figure~\ref{dangoron} illustrates how \sys operates. Each yellow block represents a pair of basic windows. Initially, we compute the exact correlation using Equation~\ref{eq:exactcorr}; if it is lower than the threshold $\beta$, it is indicated by a blue block. We then determine $k$ from the upper bound estimation, such as Equation~\ref{eq:upper_bound}, by solving for $k$ such that $Corr_{i+k-1}<\beta$ and $Corr_{i+k}\geq \beta$ using binary search. In Figure~\ref{dangoron}, $k=3$, and we use a red block to signify an upper bound higher than $\beta$. After jumping $k$ steps (skipping the green ones), we recompute the exact correlation value.

\begin{align}
\vspace{-1cm}
 \numberthis Corr_{1+k} & \leqslant  Corr_1 + \frac{1}{n_s} (k - \sum_{i}^k c_i)=Corr_{i+k}*
 \label{eq:upper_bound}
\end{align}

We have proved Equation~\ref{eq:upper_bound} under the assumption that each basic window comes from a sample distribution, as detailed in our technical report. For more complex cases, we have derived general bounds. Another feature proposed in \sys enables horizontal computation pruning. Given time-series $x$, $y$, and $z$, if we know the pairwise correlations $c_{xz}$ and $c_{yz}$, we can determine the range of $c_{xz}$ as $c_{xz}c_{yz}-\sqrt{(1-c_{xz}^2)(1-c_{yz}^2)} \leq c_{xy} \leq c_{xz}c_{yz}+\sqrt{(1-c_{xz}^2)(1-c_{yz}^2)}$. The \bm framework involves: (1) generating $\mathcal{C}$ from a user-specified distribution, (2) generating $\mathcal{x}$ in frequency space, and (3) transforming them to $\mathcal{X}$ using inverse Fourier Transform. Step (2) is based on the fact that Discrete Fourier Transform (DFT) preserves the distance between coefficients and the original time series. We developed a real-value variant of the inverse-DFT, transitioning from a complex space to a real space (unlike the original inverse-DFT, which moves from a complex to complex space), and provided proof.

\section{Results and Contributions}
\label{sec:result}
We selected TSUBASA~\cite{xu2022tsubasa} as our baseline. On the {\bf {\em NCEA Data Set}}\footnote{https://www.ncei.noaa.gov/pub/data/uscrn/products/hourly02/2020/}, \sys is an order of magnitude faster than TSUBASA in terms of pure query time and achieves an accuracy above 90 percent, comparable to Parcorr. We plan to conduct more large-scale experiments upon completing the implementation of \bm. In this work, we make the following contributions: 1) We introduce \sys, a framework that solves the problem of correlation matrix computation across sliding windows, and is at least one order of magnitude faster than the baseline; 2) We propose \bm, the first benchmark for the problem of correlation matrix computation.


\newpage

\bibliographystyle{ACM-Reference-Format}
\bibliography{ref}


\begin{thebibliography}{17}


\ifx \showCODEN    \undefined \def \showCODEN     #1{\unskip}     \fi
\ifx \showDOI      \undefined \def \showDOI       #1{#1}\fi
\ifx \showISBNx    \undefined \def \showISBNx     #1{\unskip}     \fi
\ifx \showISBNxiii \undefined \def \showISBNxiii  #1{\unskip}     \fi
\ifx \showISSN     \undefined \def \showISSN      #1{\unskip}     \fi
\ifx \showLCCN     \undefined \def \showLCCN      #1{\unskip}     \fi
\ifx \shownote     \undefined \def \shownote      #1{#1}          \fi
\ifx \showarticletitle \undefined \def \showarticletitle #1{#1}   \fi
\ifx \showURL      \undefined \def \showURL       {\relax}        \fi
\providecommand\bibfield[2]{#2}
\providecommand\bibinfo[2]{#2}
\providecommand\natexlab[1]{#1}
\providecommand\showeprint[2][]{arXiv:#2}

\bibitem[Betzel(2022)]%
        {betzel2022network}
\bibfield{author}{\bibinfo{person}{Richard~F Betzel}.}
  \bibinfo{year}{2022}\natexlab{}.
\newblock \showarticletitle{Network neuroscience and the connectomics
  revolution}.
\newblock In \bibinfo{booktitle}{\emph{Connectomic Deep Brain Stimulation}}.
  \bibinfo{publisher}{Elsevier}, \bibinfo{pages}{25--58}.
\newblock


\bibitem[Cabral et~al\mbox{.}(2017)]%
        {cabral2017functional}
\bibfield{author}{\bibinfo{person}{Joana Cabral}, \bibinfo{person}{Morten~L
  Kringelbach}, {and} \bibinfo{person}{Gustavo Deco}.}
  \bibinfo{year}{2017}\natexlab{}.
\newblock \showarticletitle{Functional connectivity dynamically evolves on
  multiple time-scales over a static structural connectome: Models and
  mechanisms}.
\newblock \bibinfo{journal}{\emph{NeuroImage}}  \bibinfo{volume}{160}
  (\bibinfo{year}{2017}), \bibinfo{pages}{84--96}.
\newblock


\bibitem[Gozolchiani et~al\mbox{.}(2008)]%
        {gozolchiani2008pattern}
\bibfield{author}{\bibinfo{person}{Avi Gozolchiani}, \bibinfo{person}{Kazuko
  Yamasaki}, \bibinfo{person}{Oz Gazit}, {and} \bibinfo{person}{Shlomo
  Havlin}.} \bibinfo{year}{2008}\natexlab{}.
\newblock \showarticletitle{Pattern of climate network blinking links follows
  El Ni{\~n}o events}.
\newblock \bibinfo{journal}{\emph{EPL (Europhysics Letters)}}
  \bibinfo{volume}{83}, \bibinfo{number}{2} (\bibinfo{year}{2008}),
  \bibinfo{pages}{28005}.
\newblock


\bibitem[Hutchison et~al\mbox{.}(2013)]%
        {hutchison2013dynamic}
\bibfield{author}{\bibinfo{person}{R~Matthew Hutchison}, \bibinfo{person}{Thilo
  Womelsdorf}, \bibinfo{person}{Elena~A Allen}, \bibinfo{person}{Peter~A
  Bandettini}, \bibinfo{person}{Vince~D Calhoun}, \bibinfo{person}{Maurizio
  Corbetta}, \bibinfo{person}{Stefania Della~Penna}, \bibinfo{person}{Jeff~H
  Duyn}, \bibinfo{person}{Gary~H Glover}, \bibinfo{person}{Javier
  Gonzalez-Castillo}, {et~al\mbox{.}}} \bibinfo{year}{2013}\natexlab{}.
\newblock \showarticletitle{Dynamic functional connectivity: promise, issues,
  and interpretations}.
\newblock \bibinfo{journal}{\emph{Neuroimage}}  \bibinfo{volume}{80}
  (\bibinfo{year}{2013}), \bibinfo{pages}{360--378}.
\newblock


\bibitem[Jayaweera and Aziz(2018)]%
        {jayaweera2018reliability}
\bibfield{author}{\bibinfo{person}{CD Jayaweera} {and} \bibinfo{person}{N
  Aziz}.} \bibinfo{year}{2018}\natexlab{}.
\newblock \showarticletitle{Reliability of principal component analysis and
  Pearson correlation coefficient, for application in artificial neural network
  model development, for water treatment plants}. In
  \bibinfo{booktitle}{\emph{IOP Conference Series: Materials Science and
  Engineering}}, Vol.~\bibinfo{volume}{458}. IOP Publishing,
  \bibinfo{pages}{012076}.
\newblock


\bibitem[Kenett et~al\mbox{.}(2010)]%
        {kenett2010dynamics}
\bibfield{author}{\bibinfo{person}{Dror~Y Kenett}, \bibinfo{person}{Yoash
  Shapira}, \bibinfo{person}{Asaf Madi}, \bibinfo{person}{Sharron
  Bransburg-Zabary}, \bibinfo{person}{Gitit Gur-Gershgoren}, {and}
  \bibinfo{person}{Eshel Ben-Jacob}.} \bibinfo{year}{2010}\natexlab{}.
\newblock \showarticletitle{Dynamics of stock market correlations.}
\newblock \bibinfo{journal}{\emph{AUCO Czech Economic Review}}
  \bibinfo{volume}{4}, \bibinfo{number}{3} (\bibinfo{year}{2010}).
\newblock


\bibitem[Kim et~al\mbox{.}(2019)]%
        {kim2019complex}
\bibfield{author}{\bibinfo{person}{Kyunghun Kim}, \bibinfo{person}{Hongjun
  Joo}, \bibinfo{person}{Daegun Han}, \bibinfo{person}{Soojun Kim},
  \bibinfo{person}{Taewoo Lee}, {and} \bibinfo{person}{Hung~Soo Kim}.}
  \bibinfo{year}{2019}\natexlab{}.
\newblock \showarticletitle{On complex network construction of rain gauge
  stations considering nonlinearity of observed daily rainfall data}.
\newblock \bibinfo{journal}{\emph{Water}} \bibinfo{volume}{11},
  \bibinfo{number}{8} (\bibinfo{year}{2019}), \bibinfo{pages}{1578}.
\newblock


\bibitem[Krendl and Betzel(2022)]%
        {krendl2022social}
\bibfield{author}{\bibinfo{person}{Anne~C Krendl} {and}
  \bibinfo{person}{Richard~F Betzel}.} \bibinfo{year}{2022}\natexlab{}.
\newblock \showarticletitle{Social cognitive network neuroscience}.
\newblock \bibinfo{journal}{\emph{Social Cognitive and Affective Neuroscience}}
  \bibinfo{volume}{17}, \bibinfo{number}{5} (\bibinfo{year}{2022}),
  \bibinfo{pages}{510--529}.
\newblock


\bibitem[Mueen et~al\mbox{.}(2010)]%
        {mueen2010fast}
\bibfield{author}{\bibinfo{person}{Abdullah Mueen}, \bibinfo{person}{Suman
  Nath}, {and} \bibinfo{person}{Jie Liu}.} \bibinfo{year}{2010}\natexlab{}.
\newblock \showarticletitle{Fast approximate correlation for massive
  time-series data}. In \bibinfo{booktitle}{\emph{Proceedings of the 2010 ACM
  SIGMOD International Conference on Management of data}}.
  \bibinfo{pages}{171--182}.
\newblock


\bibitem[Sakurai et~al\mbox{.}(2005)]%
        {sakurai2005braid}
\bibfield{author}{\bibinfo{person}{Yasushi Sakurai}, \bibinfo{person}{Spiros
  Papadimitriou}, {and} \bibinfo{person}{Christos Faloutsos}.}
  \bibinfo{year}{2005}\natexlab{}.
\newblock \showarticletitle{Braid: Stream mining through group lag
  correlations}. In \bibinfo{booktitle}{\emph{Proceedings of the 2005 ACM
  SIGMOD international conference on Management of data}}.
  \bibinfo{pages}{599--610}.
\newblock


\bibitem[Thirion et~al\mbox{.}(2014)]%
        {thirion2014fmri}
\bibfield{author}{\bibinfo{person}{Bertrand Thirion}, \bibinfo{person}{Ga{\"e}l
  Varoquaux}, \bibinfo{person}{Elvis Dohmatob}, {and}
  \bibinfo{person}{Jean-Baptiste Poline}.} \bibinfo{year}{2014}\natexlab{}.
\newblock \showarticletitle{Which fMRI clustering gives good brain
  parcellations?}
\newblock \bibinfo{journal}{\emph{Frontiers in neuroscience}}
  \bibinfo{volume}{8} (\bibinfo{year}{2014}), \bibinfo{pages}{167}.
\newblock


\bibitem[Tilfani et~al\mbox{.}(2021)]%
        {tilfani2021dynamic}
\bibfield{author}{\bibinfo{person}{Oussama Tilfani}, \bibinfo{person}{Paulo
  Ferreira}, \bibinfo{person}{El Boukfaoui}, {and} \bibinfo{person}{My
  Youssef}.} \bibinfo{year}{2021}\natexlab{}.
\newblock \showarticletitle{Dynamic cross-correlation and dynamic contagion of
  stock markets: A sliding windows approach with the DCCA correlation
  coefficient}.
\newblock \bibinfo{journal}{\emph{Empirical Economics}} \bibinfo{volume}{60},
  \bibinfo{number}{3} (\bibinfo{year}{2021}), \bibinfo{pages}{1127--1156}.
\newblock


\bibitem[Xu et~al\mbox{.}(2022)]%
        {xu2022tsubasa}
\bibfield{author}{\bibinfo{person}{Yunlong Xu}, \bibinfo{person}{Jinshu Liu},
  {and} \bibinfo{person}{Fatemeh Nargesian}.} \bibinfo{year}{2022}\natexlab{}.
\newblock \showarticletitle{TSUBASA: Climate Network Construction on Historical
  and Real-Time Data}. In \bibinfo{booktitle}{\emph{Proceedings of the 2022
  International Conference on Management of Data}}. \bibinfo{pages}{286--295}.
\newblock


\bibitem[Yagoubi et~al\mbox{.}(2018)]%
        {yagoubi2018parcorr}
\bibfield{author}{\bibinfo{person}{Djamel~Edine Yagoubi}, \bibinfo{person}{Reza
  Akbarinia}, \bibinfo{person}{Boyan Kolev}, \bibinfo{person}{Oleksandra
  Levchenko}, \bibinfo{person}{Florent Masseglia}, \bibinfo{person}{Patrick
  Valduriez}, {and} \bibinfo{person}{Dennis Shasha}.}
  \bibinfo{year}{2018}\natexlab{}.
\newblock \showarticletitle{ParCorr: efficient parallel methods to identify
  similar time series pairs across sliding windows}.
\newblock \bibinfo{journal}{\emph{Data Mining and Knowledge Discovery}}
  \bibinfo{volume}{32} (\bibinfo{year}{2018}), \bibinfo{pages}{1481--1507}.
\newblock


\bibitem[Zhang et~al\mbox{.}(2009)]%
        {zhang2009adaptive}
\bibfield{author}{\bibinfo{person}{Tiancheng Zhang}, \bibinfo{person}{Dejun
  Yue}, \bibinfo{person}{Yu Gu}, \bibinfo{person}{Yi Wang}, {and}
  \bibinfo{person}{Ge Yu}.} \bibinfo{year}{2009}\natexlab{}.
\newblock \showarticletitle{Adaptive correlation analysis in stream time series
  with sliding windows}.
\newblock \bibinfo{journal}{\emph{Computers \& Mathematics with Applications}}
  \bibinfo{volume}{57}, \bibinfo{number}{6} (\bibinfo{year}{2009}),
  \bibinfo{pages}{937--948}.
\newblock


\bibitem[Zhong et~al\mbox{.}(2020)]%
        {zhong2020filcorr}
\bibfield{author}{\bibinfo{person}{Sheng Zhong}, \bibinfo{person}{Vinicius~MA
  Souza}, {and} \bibinfo{person}{Abdullah Mueen}.}
  \bibinfo{year}{2020}\natexlab{}.
\newblock \showarticletitle{FilCorr: Filtered and lagged correlation on
  streaming time series}. In \bibinfo{booktitle}{\emph{2020 IEEE International
  Conference on Data Mining (ICDM)}}. IEEE, \bibinfo{pages}{1436--1441}.
\newblock


\bibitem[Zhu and Shasha(2002)]%
        {zhu2002statstream}
\bibfield{author}{\bibinfo{person}{Yunyue Zhu} {and} \bibinfo{person}{Dennis
  Shasha}.} \bibinfo{year}{2002}\natexlab{}.
\newblock \showarticletitle{Statstream: Statistical monitoring of thousands of
  data streams in real time}. In \bibinfo{booktitle}{\emph{VLDB'02: Proceedings
  of the 28th International Conference on Very Large Databases}}. Elsevier,
  \bibinfo{pages}{358--369}.
\newblock


\end{thebibliography}

\end{document}